\documentclass[journal]{IEEEtran}
\usepackage[cmex10]{amsmath}
\usepackage{amsfonts,amsthm,amssymb}
\usepackage{mathtools}
\usepackage{xcolor}
\usepackage[T1]{fontenc}
\usepackage[font=scriptsize,caption=false,labelsep=space]{subfig}
\usepackage{graphicx,float,wrapfig}
\usepackage{upquote}
\usepackage[none]{hyphenat}
\usepackage{url}
\usepackage{upgreek}
\usepackage{cite}
\usepackage{glossaries}
\usepackage[named]{algo}
\usepackage[noend]{algpseudocode}
\usepackage{algorithm}
\graphicspath{{./figures/}}

\newacronym{SNR}{SNR}{Signal to Noise Ratio}
\newacronym{INR}{INR}{Interfernece to Noise Ratio}
\newacronym{SINR}{SINR}{Signal to Interference plus Noise Ratio}
\newacronym{MIMO}{MIMO}{Multiple Input Multiple Output}
\newacronym{SISO}{SISO}{Single Input Single Output}
\newacronym{MM}{MM}{Majorization-Minimization}
\newacronym{DFT}{DFT}{Discrete Fourier Transform}
\newacronym{FFT}{FFT}{Fast Fourier Transform}
\newacronym{BPSK}{BPSK}{Binary Phase Shift Keying}
\newacronym{QPSK}{QPSK}{Quadrature Phase Shift Keying}
\newacronym{ULA}{ULA}{Uniform Linear Array}
\newacronym{DOF}{DOF}{Degrees of Freedom}
\newacronym{PSK}{PSK}{Phase Shift Keying}
\newacronym{MPSK}{MPSK}{M-ary Phase Shift Keying}
\begin{document}
\title{Precoding for Satellite Communications: Why, How and What next?}
\author{Bhavani Shankar Mysore. Eva Lagunas. Symeon Chatzinotas, Bjorn Ottersten. % <-this % stops a space
\thanks{The authors are with the SnT - Interdisciplinary Centre for Security, Reliability and Trust, University of Luxembourg,
Luxembourg, e-mail: \{bhavani.shankar, eva.lagunas, symeon.chatzinotas, bjorn.ottersten\}@uni.lu.}
\thanks{%This work was supported by FNR (Luxembourg) through the BRIDGES project ``AWARDS'', C-PPP17/IS/11827256/AWARDS, and CORE project ``SPRINGER'', C18/IS/12734677/SPRINGER. 
This work was supported by Luxembourg National Research Fund (FNR) through the projects PROSAT, DISBUS.}
\vspace*{-0.22in}
}
\maketitle
\begin{abstract}
Precoding has stood out as a promising multi-user transmission technique to meet the emerging throughput demand of satellite communication systems while awaiting the technological maturity for exploiting higher bands. Precoding enables the reduction of  interference among co-channel beams through spatial processing while promoting aggressive frequency reuse and improving spectral efficiency. Satellite systems offer multitude of system and service configurations, resulting in different precoder design methodologies. This article explores the motivation for the introduction of precoding, offers an insight to their theoretical development in a diverse scenarios and presents some avenues for future development.
 \end{abstract}
%
%%------------------------------------
%% The introduction starts here
%%------------------------------------
%
\vspace*{-0.2in}
\section{Introduction}
The reinvention of satellite systems towards offering broadband services, planned integration with 5G both for access and backhaul and the scarcity in traditional frequency bands has motivated the actors in satellite systems to seek alternatives. A natural way forward is to move to higher bands, like Q/ V/ W bands. This approach takes time due to the large investment needed for infrastructure building and the necessity of devising techniques and technologies mature enough to offer robust technical solutions against the impairments at higher bands. Another approach is to reuse the existing spectrum bands with even higher efficiency. Several interference management techniques have been considered in literature and an interesting technique that blends well with the ubiquitous multibeam satellites is precoding \cite{spm}. 

Downlink precoding techniques have been widely studied in multi-antenna cellular communication systems to overcome co-channel interference
(CCI) introduced by frequency reuse \cite{DL_Prec}. Following the terrestrial trends,  multiple spot beam satellites reuse available bandwidth among adjacent beams and  precoding was proposed to mitigate inter-beam interference (IBI) \cite{Devillers2011}, \cite{Taricco2014}, \cite{Vazquez2016}. By exploiting the channel state information (CSI) of the user terminals at the gateway (GW), these works aimed at devising a linear precoder for simultaneous transmission to multiple users in an unicast manner. Several seminal works have been since pursued investigations on improving the precoder design, extensions to multicasting and on-board implementation, as well as low complexity designs robust to channel and system imperfections including the non-linearities.  Further, precoding is now supported in terms of framing and signaling in the latest  DVB-S2X  standard; see in particular \cite{DVB-S2X}.  The industry has also shown interest with a planned live demonstration of  precoding over the satellite \cite{kodheli2020satellite}.

This work  presents a canvas of developments in precoding for satellite systems.  Starting from the typical unicast scenario, it first  explores the increasingly complex precoding formulations for different scenarios culminating in the satellite oriented frame-based precoding. Ideal conditions are assumed and the focus will be on Signal to Interference plus Noise Ratio (SINR) optimization;  methodologies and constraints used therein are highlighted. Subsequently, the consideration will be on incorporating practical aspects including imperfections and feasibility of implementation. In this context, robust feasible precoding solutions are discussed.  Interesting research avenues will be presented towards addressing challenges in the evolving satellite ecosystem.  To present this canvas,  the paper is organized as follows: Section II presents the system model and Section III discusses the precoding techniques for different satellite scenarios under ideal conditions.  Section IV focusses on robust designs considering practical design problems. Finally,  the future directions and open challenges are presented in Section V.
\vspace*{-0.1in}
\section{Multibeam Satellite System}
\label{sec:Prec_ideal}
A generic multibeam satellite system having $N_b$ fixed beams generated from $N_f$ feeds is considered. Without loss of generality, each beam is assumed to serve an identical $N_u$ number of users.  The way these users are served leads to different scenarios that will be discussed later. Normally adjacent beams transmit in different frequencies/ polarization to avoid IBI; herein we consider identical frequency/ polarization on all the beams (also termed full frequency reuse). In this context, beam generation becomes central towards reducing IBI; this process typically involves,

{\sl 1) Antenna section:}  It includes the feed and reflector assembly in one of the configurations: (i) each feed generating a beam using reflector ($N_f=N_b$), (ii) multiple feeds creating the beam using a reflector ($N_f\geq N_b$) and (iii) direct radiating array creating beams without reflectors ($N_f\geq N_b$).

{\sl 2) Beamforming Network (BFN):} This processing element transforms the signals intended for the beams to those being transmitted from the feeds.  In the context of full frequency reuse, BFN is essential to avoid IBI with minimal processing at the user. Typically, the BFN is a linear combiner whose weights determine the beam properties and IBI mitigation. Such a   BFN  is typically implemented on-ground at the GWs, while the new generation also enables an hybrid on-board (the satellite)/ on-ground implementations.
%\begin{itemize}
 %   \item Antenna section: This includes feeds and reflector used for creating the needed electromagnetic pattern. It includes, (i) single feed generating the beam using reflector, (ii) multiple feeds creating the beam using a reflector and (iii) direct radiating array (DRA) creating beams without reflectors. 
%
  %   \item Beamforming Network (BFN): This processing element transforms the signals intended for the beams to those being transmitted from the feeds.  In the context of full frequency reuse, BFN is essential to avoid IBI with minimal processing at the user. Typically, the BFN is a linear combiner whose weights determine the beam properties and IBI mitigation. Such a   BFN  is typically implemented on-ground at the GWs, while the new generation also enables an hybrid on-board (the satellite)/ on-ground implementations.
%
%\end{itemize}
%In this case, the  $N_b$ beams are generated by the linear combination of the transmissions from $N_f$  feeds (antennas) on-board the satellite through a reflector. The weights determining the linear combination leads to   

The dynamics of the BFN elements  leads to the two often confused paradigms in satellite communications $-$ beamforming and precoding.    This is clarified first.
%The nature and location of the BFN leads to different paradigms.  The dynamics of the weights used in the linear combination
\subsection{Beamforming and Precoding}
\label{ssec:Prec_BF}
Typically, the multiple beams on the satellite are formed to offer certain coverage on the ground based on the traffic requirements. A classical example is country specific beams offering particular language content. Often, the beams are also shaped to avoid too much discrepancy in traffic demands within the coverage; this helps in efficient resource utilization. Such beam designs are quasi-static and vary in a long-term in response to changes in the temporally averaged traffic. The beams formed in this context do not exploit the instantaneous CSI of the users and hence are not optimized for the individual user; however, they are designed considering propagation conditions and requirements of all the users in the field of view/ coverage area. This  user-agnostic, traffic-aware and quasi-static BFN paradigm leads to {\em beamforming}. 

On the contrary, {\em precoding} refers to the BFN exploiting the instantaneous CSI\footnote{here, while CSI refers to the end-to-end path from the GW to the users through the satellite, the GW to satellite channel is typically assumed fixed and known. Hence CSI specializes to downlink channel from satellite and is obtained through explicit feedback from users.} to a group of users.  This leads to optimized transmissions to the selected set of users. Precoding offers a traffic and coverage agnostic, CSI dependent transmission.  Since CSI changes, the BFN needs to be adaptive. 

Depending on the payload processing and feeder link (GW to satellite) constraints, these techniques can be implemented on-board the satellite or on-ground at GW. The  BFN is typically implemented on-board using analog components and reflector shaping due to limited variability and is also known as analog beamforming. On the other hand, precoding is implemented on-ground using base-band digital processing. On-ground implementation requires larger feeder link resources since the feed signals (and not beam signals) need to be uplinked. Typically, the two techniques are designed separately and the focus here is  on precoding; however, the paper presents a scenario later on their joint design.

%
%In a multibeam satellite, an on-board quasi-static beamforming and an on-ground dynamic precoding operation are present. %
%Focussing on a particular time instance, let ${\bf h}^k_i$ be the $N_f\times 1$ channel vector from the $N_f$ feeds to the $i$th single antenna UT in the $k$th beam. This channel comprises link-budget parameters including antenna gains, signal attenuation due path-loss,  back-off and antenna pointing loss. Additionally, the channel also includes the impact of radiowave propagation, thereby incorporating small scale fading and rain-fading at higher bands. Different channel models have been considered for modelling the fading [SPM]. Let the $i$th UT in the $k$th beam be interested in the symbol $s_i^k$. The precoding operation and the signal received by this UT due to full-frequency reuse follows as, 
%
\section{Precoding: From Unicast to Frame-based}
%
%\subsubsection{Precoding Model}
%\label{sssec:Prec}
%
%Focussing on a particular time instance, let ${\bf h}_k$ be the $N_f\times 1$ channel vector from the $N_f$ feeds to a generic $k$th single antenna user. This channel comprises link-budget parameters including antenna gains, signal attenuation due path-loss,  back-off and antenna pointing loss. Additionally, the channel also  incorporates small scale fading and rain-fading at higher bands. Different channel models have been considered for modelling the fading \cite{kodheli2020satellite}.
%
Let $s_i$ be distinct communication symbols needed to be transmitted on the $i$th beam, $,~i\in [1, N_b]$. The $N_f\times 1$ precoding vector, ${\bf w}_i$, transforms the beam signal $s_i$ to the feed transmissions, ${\bf w}_i s_i$. Further, let  the users in the $i$th beam interested in $s_i$ be denoted by  ${\cal U}_i$. Focussing on a particular time instance, let ${\bf h}_k$ be the $N_f\times 1$ channel vector from the $N_f$ feeds to a generic $k$th single antenna user. This channel comprises link-budget parameters including antenna gains, signal attenuation due path-loss,  back-off and antenna pointing loss. Additionally, the channel also  incorporates small scale fading and rain-fading at higher bands; kindly refer to \cite{kodheli2020satellite} for the different channel models. The choice of the users affects the channels and hence the CSI dependent precoding vector; thus  it is essential to qualify the sets ${\cal U}_i$; this is pursued next.
\vspace*{-0.1in}
\subsection{Unicast scenario}
\label{ssec:Uni}
In the unicast scenario, the set ${\cal U}_i$ contains only one user per beam and can be realized by employing time division multiple access (TDMA) for users in each beam. Thus, the design assumes $N_b$ users, with the user index also denoting the associated beam. Further, $N_b$ precoding vectors, $\{{\bf w}_i\}$, are used to optimize the system. Considering the $i$th user (in beam $i$),  the corresponding received signal, $y_{i}$, and the resulting SINR are given by,
% it follows that,
%
\begin{eqnarray}
\label{eq:gen_prec}
y_{i} &=& {\bf h}_{i}^H {\bf w}_i s_i  + \sum_{j\in {\cal U}, j\neq i} {\bf h}_i^H {\bf w}_j s_j + n_i, ~~~~ i\in {\cal U}, \\
\label{eq:SINR_gen}
{\rm SINR}_i &=&\frac {|{\bf h}_i^H {\bf w}_i|^2 }{\sum_{j=1, j\neq i} |{\bf h}_i^H {\bf w}_j|^2 + \sigma^2_i}, ~~~~ i\in [1, N_b],
\end{eqnarray}
where  $s_j, {\bf w}_j$ are the data and precoding vectors for user $j$ respectively and  $n_i$ is the additive zero-mean white Gaussian noise with variance $\sigma_i^2$. Further, $\{s_k\}$ are assumed independent with zero mean and unit variance. The aim is to design the precoders to enhance system performance and  classical approaches include,
\vspace*{0.07in}
%\begin{table}[t]
%\label{tab:Uni}
%    \caption{Canvas of objective functions and constraints for Precoder design}
 %   \centering
%

\begin{tabular}{lr} 
$
      \begin{array}{c}
        {\rm Power~~Minimization}  \vspace*{0.08in} \\ \hline 
         \min_{ \{{\bf w}_i\} } ~~ \sum_{i=1}^{N_b} || {\bf w}_i||^2 \vspace*{0.08in}  \\
        {\rm s.~t}~~ {\rm SINR}_i \geq \gamma_i 
      \end{array}
      $
      &
$    \begin{array}{c}
  {\rm Max-min~fair}  \vspace*{0.08in}  \\ \hline 
   \max_{\{{\bf w}_i\}} \min_{i\in N_b}  {\rm SINR}_i \vspace*{0.08in} \\
  {\rm s.~t~~~} \sum_{i=1}^{N_b} || {\bf w}_i||^2\leq P_T 
     \end{array}$
     \vspace*{0.08in}
\end{tabular}
where $\gamma_i$ is the threshold for user $i$ to satisfy certain rate constraints. $||\cdot||$ is the Euclidean norm and $P_T$ is the total transmit power among all the beams arising from the use of multi-port amplifiers. In the max-min fair problem above, a total power constraint is considered. Alternatively,  satellite systems also need a per-feed power constraint since each feed typically has its own power source. The per-antenna power constraint takes the form $\sum_{i} \left[ {\bf w}{\bf w}^H \right]_{i, i} \leq P_i$, where $\left[A\right]_{k, k}$ is the $(k, k)$ diagonal entry of matrix ${\bf A}$, $P_i$ is the power of $i$th feed and $^H$ denotes Hermitian operation. The other related problem is to maximize the sum rate $\sum_{i=1}^{N_b} \log_2 (1+{\rm SINR}_i)$ subject to the power constraints.

Optimal precoder solutions depend on the problem formulation in general. For power minimization problems with ${\rm SINR}$ constraints, the classical Semi-definite relaxation (SDR) method was proposed \cite{Mats}. Second order cone programming (SOCP) has also been applied to exploit hidden convexity. The {\em max-min fair} problem  was solved using duality, which also offers elegant framework for solving the power minimization problem \cite{stanczak2009fundamentals}. The framework was later extended to the per-antenna power constraint \cite{christopoulos2014weighted}. On the other hand,  simple linear precoders include the Zero-Forcing (ZF) and the minimum Mean square error (MMSE); particularly, the latter, offers a good trade-off between performance and complexity. The precoder design for unicast scenario is mature and is worth revisiting when novel system constraints arise. 
 
In the above formulations, the user associated with ${\cal U}_i$ was already identified. However, in satellite systems, there are usually more than one user in a beam needing service. In the unicast scenario, this necessitates some sort of an user-selection or {\em scheduling} where the sets ${\cal U}_i$ are designed  to include  users who (i) have good channel conditions and (ii) offer limited interference to others. Naturally, the optimal transmit scheme involves design of the inter-dependent scheduling and precoding algorithms.  Several ad-hoc and approximate optimization approaches have been proposed for this NP-hard problem \cite{bandi2019joint}. A representative joint design selecting $N_b$ users among a total of $N_b N_u$ users and maximizing a weighted sum-rate over the continuous variables, i. e., precoding vectors, and the binary scheduling variables, i. e.,  $\eta_i$ indicating the selection of $i$th user, is given below.
\vspace*{-0.08in}
  \begin{align}\label{eq:Bin_WSR_prob}
    {\cal P}_1 &: \text{ } \max_{\{{\bf w}_i\},\{P_i\}, \{\eta_i\}}  \sum_{i=1}^{N_u N_b}  \beta_{i} \log_2 (1+{\rm SINR}_i)
    \\ 
    \rm{s.~t,~} &  \eta_{i} \in \lbrace 0,1 \rbrace, \text{ } ||{\bf w}_{i}||^2\leq \eta_{i} \overline{P_i}, \sum_{i=1}^{N_u N_b} \eta_{i} \leq N_b, \sum_{i=1}^{N_u N_b} \overline{P_i} \leq P_T,  \nonumber \end{align}
where $\overline{P_i}$ is the power of the $i$th precoder, $\eta_i=1$ if $i$th user among the total pool of $N_u N_b$ users is scheduled (and zero otherwise), and $\beta_i$ are the system specific weights \cite{bandi2019joint}.
%Note that \eqref{eq:Bin_WSR_prob} assumes $N_
%
\vspace*{-0.08in}
\subsection{Multicast scenario}
\label{ssec:Mult}
%
%In the unicast setting, each user was served with an independent data stream. The other extreme is the broadcast of information to all the users. 
%In many situations, a scenario in between the two presented above arises: only a set of users requiring access to same data. %
Broadcasting common data to all users in the coverage is the other extreme of scheduling. In many cases, a hybrid unicast and broadcast scenario arises where only a set of users require access to common data. A case in point could be streaming of a local event to a small set of population; broadcasting to each population group needs appropriate beamforming requiring CSI. In this context, multicasting avoids the resource wastage arising from transmitting the same message over different resources to the users and has become part of the new generation of communication standards. Several variants like physical layer multicasting (where multiple users need common data) or multigroup multicasting (users requiring common data are grouped) exist; in this work, these variants are dealt  under a generic {\em multicast} scenario.
%In the initial works, the concept of physical layer multicasting was proposed where a common information needs to be broadcast to a set of users whose CSI was known at the transmitter. This was later extended to multigroup multicasting where users were split into multiple groups with users in each group requiring a common message. In this work, these variants are dealt under {\em multicast scenario}.
%Another embodiment of this multicast applied in satellite communications is the Frame-based precoding; this will be treated in Section \ref{sssec:Frame}. 

The signal model  in \eqref{eq:gen_prec}  can be generalized to the multicast scenario. Noting that users requiring common data do not interfere with each other, the ${\rm SINR}_i$ takes the form,
%and the re In the considered scenario, the set ${\cal U}=\bigcup_{i=1}^{N_b} {\cal U}_i$, where each ${\cal U}_i$ contains users associated with the $i$th beam. However, all the users in ${\cal U}_i$ require the data $s_i$. Since there are only $N_b$ independent data-streams, there only exist $N_b$ independent precoders. Further, users desirous of common information do not cause interference to each other. The ${\rm SINR}$ for UT $j$ in beam $k$, takes the form,
%
\begin{eqnarray}
%\label{eq:gen_prec}
%y_{i} &=& {\bf h}_{i}^H {\bf w}_i s_i  + \sum_{j\in {\cal U}, j\neq i} {\bf h}_i^H {\bf w}_j s_j + n_i, ~~~~ i\in {\cal U}, \\
%
\label{eq:mult_prec}
{\rm SINR}_{i} &=& \frac {|{\bf h}_{i}^H {\bf w}_k|^2 }{\sum_{j=1, j\neq k}^{N_b}  |{\bf h}_{i}^H {\bf w}_j|^2 + \sigma^2_i}, ~i\in {\cal U}_k, ~k \in[1, N_b].
\end{eqnarray}
For a given user grouping, $\{ {\cal U}_i\}$, this ${\rm SINR}_\ast$ can be used in any of the precoding design problems  mentioned earlier.  However, the following aspects need to be noted,
\begin{itemize}
    \item Limited degrees of freedom: Only $N_b$ precoders serving more than $N_b$ users simultaneously. 
    \item Increased requirements: Each user requires particular ${\rm SINR}$ to stay connected.
\end{itemize}
%
%Thus the number of design variables are much less than the requirements. This makes the optimization problems harder. When the user association to the groups is already provided, the optimization problem can take any of the objective/ constraint combinations provided in Table \ref{tab:Uni}, but the number of ${\rm SINR}$ variables, while still being the cardinality of ${\cal U}$, is much higher than $N_b$.
The optimal precoder design for multicast scenario is NP-hard and several approaches have been pursued towards approximately optimal solutions. Many of the early works consider a SDR based approach for power minimization  and  max-min fair problem under total power constraints \cite{sidiropoulos2006transmit}, \cite{karipidis2008quality}. The SDR approach was extended to the case of per-antenna constraints in \cite{christopoulos2014weighted}. Letting, ${\bf Q}_k={\bf h}_k{\bf h}^H_k, {\bf W}_k={\bf w}_k{\bf w}^H_k$, $Tr()$ to be the trace operator and recalling $P_i$ to be the $i$th feed power,  the SDR approach casts the max-min fair problem as, 
  \begin{align}\label{eq:SDR}
    {\cal P}_2 &: \text{ } \max_{\{{\bf W}_k\}} t
    \\ 
    \rm{s.~t~},& {\rm SINR}_i =\frac{Tr({\bf Q}_i{\bf W}_k)}{\sum_{j\neq k} Tr({\bf Q}_i{\bf W}_j)   + \sigma^2_i} \geq t,  {\bf W}_k \succeq {\bf 0},  \nonumber\\
   &  \sum_{j=1}^{N_b}[{\bf W}_j]_{l, l} \leq P_l,~ i\in {\cal U}_k, ~l, k \in[1, N_b],
    \end{align}
and finds the rank 1 approximation to optimal $\{{\bf W}_k\}$. 

To overcome the scalability issues with the SDR based method, iterative approaches exploiting the quadratically constrained quadratic program (QCQP) nature as well as difference of convex functions based methodologies have been proposed \cite{gharanjik2016max}, \cite{bandi2019joint}. A   QCQP formulation for the power minimization problem takes the form \cite{gharanjik2016max},  
  \begin{align}\label{eq:QCQP}
    {\cal P}_3 &: \text{ }\text{ } \max_{ {\bf w}  }   {\bf w}^H {\bf w}
    \\ 
    \rm{s.~t,~}&  {\bf w}^H {\bf R}_j{\bf w},~~j\in {\cal U}_k, ~~ k \in[1, N_b] \nonumber,  \end{align}
where ${\bf w}=[{\bf w}^H_1, {\bf w}_2^H, \ldots, {\bf w}_{N_b}^H]^H$, ${\bf R}_j=\frac{1}{\sigma_i^2 \gamma_i}\left(\widehat{\bf R}_i-\gamma_i \widetilde{\bf R}_i\right)$, $\widehat{\bf R}_i= diag({\bf e}_i)\otimes {\bf h}_i{\bf h}_i^H, \widetilde{\bf R}_i= \left({\bf I}-diag({\bf e}_i)\right)\otimes {\bf h}_i{\bf h}_i^H $ with ${\bf e}_k$ being a $N_b$ dimensional $k$th standard basis vector, $\otimes$ is the Kronecker product and ${\bf I}$ is a $N_b$ dimensional identity matrix. 

As in unicast, the optimal system design would also involve the selection of each of the ${\cal U}_i$. Issues that were discussed in the unicast case also arise here, but the scheduling problem is accentuated by the increased number of users. In addition, in the most general case, each beam has a large number of user groups and not just limited to one. Thus in addition to the scheduling of users within the groups, another round of {\em group-scheduling} needs to be pursued. The emergence of satellite communications has rekindled the research interest in joint scheduling and precoder design for multicast scenario \cite{bandi2019joint}. Since the original problem in NP-hard, obtaining low complexity efficient precoder designs for systems with large dimensions and novel constraints is an active area of research. 
\subsubsection{Frame based precoding}
\label{sssec:Frame}
Frame-based precoding is a variant of multicasting arising due to the frame structure used in multibeam satellite communications. In the traditional noise-limited and fade-prone satellite communications, long forward error correction (FEC) codes were used to enhance the link-budget. To avoid resource wastage through dummy frames, data from different users in a beam are multiplexed into each of these long FEC codewords. In a multibeam system with aggressive frequency reuse, precoding is undertaken on data streams corresponding to co-channel beams.  While the new standard DVB-S2x supports precoding, it does so on a frame-by-frame basis; this precludes user-by-user precoding. In this context, the users in a beam whose data is multiplexed, need to decode the whole frame to extract relevant information. Thus, the situation can be construed as multicasting.  %composite frame is then transmitted to ensure decoding by the user with lowest ${\rm SINR}$.

In the context of frame based precoding, the design objectives and constraints are similar to those in mentioned under the multicast scenario; the exception being the use of spectral efficiency offered by the modulation and coding schemes of DVB-S2x instead of the Shannon rate \cite{christopoulos2015multicast}. Many of these problems have been pursued in \cite{christopoulos2015multicast} following an iterative optimization of user scheduling and precoding.
\section{Robust Designs for Precoding}
\label{ssec:Robust}
In the earlier section, an ideal linear channel with perfect CSI is assumed. However, in a satellite system, many of the following non-idealities arise warranting a study of robust precoder designs:
%
%\begin{itemize}
%\item 
\paragraph{Imperfect CSI} Precoding is typically performed at the GW. However, the CSI used for this design is outdated due to long round trip time  (RTT) of about 250 ms in each direction. In this interval, the downlink radio-wave channel can vary due to movement of the user or of scatterers. This causes outdated CSI at GW which impacts precoding and the selection of appropriate modulation and coding \cite{precoding-scheduling-and-link-adaptation-in-mobile-interactive-multibeam-satellite-systems}. Further, the phase noise process of the satellite transponder, which is absorbed into the CSI, varies during RTT. This introduces time-varying differential phase shifts to CSI, thereby impacting its quality. 
%
%\item
\paragraph{Non-linearities} The satellite transponder includes the high power amplifier (HPA). whose operation is inherently non-linear. Further many of the analog components like mixers tend to introduce inter-modulation products. This destroys the attractive linear model of \eqref{eq:gen_prec}, forcing to rethink precoder designs.
%\end{itemize}

In the following, two such avenues will be discussed.
\vspace*{-0.15in}
\subsection{Designs coping with Phase Noise} 
\label{ssec:PN}
Considering transmissions to fixed users, one may recognize that rain attenuation is the key dynamic component affecting the channel amplitude. Such variations are typically slow and  the channel amplitude can be assumed  fixed during the RTT. On the other hand, there is a significant variation in the channel phase arising from the different time-varying phase components with phase noise of the on-board local oscillator (LO) being
the dominant contributor. Thus, the dynamic nature of the channel and a high RTT lead to outdated estimates of the channel phase. Therefore, the performance of the system becomes unpredictable when the GW uses outdated CSI due to phase uncertainty.

Let the actual channel be ${\bf h}_i$, while the estimated channel, used for precoder optimization, be ${\bf h}_i \odot e^{j{\bf \Theta}_i}$, where ${\bf \Theta}_i$ is the vector phase noise process. Herein, each feed is assumed to have its own transponder.  Several distributions including uniform, Tikhonov and Gaussian distributions are ascribed to this process, each exploiting the apriori knowledge (or lack thereof) and for a particular region of operation \cite{martinez2019effects}.  Precoder designs  that are robust under different criteria have been formulated as optimization problems and solved in \cite{gharanjik2015robust}. The ${\rm SINR}_k$ is now a random variable, and related criteria include, %
\begin{tabular}{lr} 
%\vspace*{0.1in}
\hspace*{-0.15in}
$
      \begin{array}{l}
        {\rm Outage~~ Minimization}  \vspace*{0.08in} \\ \hline \min_{ \{{\bf w}_i\} }\max_k  {\rm Prob} ({\rm SINR}_k\leq \epsilon)
          \end{array}
      $
      &
$    \begin{array}{l}
        {\rm Average~~ SINR}  \vspace*{0.08in} \\ \hline \max_{ \{{\bf w}_i, k\} } {\rm E} ({\rm SINR}_k)
     \end{array}$
%     \vspace*{0.08in}
\end{tabular}
impact  on the system level is shown in \cite{martinez2019effects}. These works indicate that peculiarities of the satellite system warrant a careful analysis of components that are often neglected.  
% SOMETHING MORE?
%
\vspace*{-0.1in}
\subsection{Designs coping with Non-linearity} 
\label{ssec:NL}
Most of the works on precoding mitigate the linear co-channel
interference between the beams caused by frequency reuse. However, the HPA, an integral part of the satellite payload, is inherently non-linear. Non-linear amplification combined with the linear co-channel interference introduces non-linear co-channel distortions at the receiver. Further, signals with very high peak to average power ratios (PAPR), typical of spectrally efficient modulations like 16/ 32 point multi-ring constellations, are sensitive to the non-linear characteristic of the HPA and necessitate large back-off to have manageable
distortion levels; large back-off naturally reduces power amplification efficiency and the useful signal power.

To understand the impact of non-linearity, a first step is its modelling. Several models exist; a simple third order memory-less model based on Volterra series for the received signal of the $i$th user takes the form \cite{mengali2016joint}, 
\begin{eqnarray}
\label{eq:Prec_def}
&{\bf x} = {\bf W}{\bf s}, ~~{\bf s}=[s_1, \ldots, s_{N_b}]^T, ~~{\bf W}=[{\bf w}_1, \ldots, {\bf w}_{N_b}]^T,  \\
\label{eq:non_linear}
&y_i = g_{1, i}{\bf h}_i^H  {\bf x} + g_{3, i} {\bf h}_i^H  \left( {\bf x}\odot {\bf x} \odot {\bf x}^\ast\right) +  n_i, ~~~~ i\in {\cal U},&
\end{eqnarray}
where $\odot$ is the Hadamard product, $^\ast$ is the complex conjugate, $g_{1, i}$ and $g_{3, i}$ are the first and third order Volterra coefficients. It follows from \eqref{eq:non_linear}, that the ${\rm SINR}$ is a non-linear function of the precoding matrix ${\bf W}$. The earlier presented works are not applicable in this scenario and different signal processing approaches need to be pursued. 

One approach is to impose conditions on the maximum signal amplitude, variously known as crest factor reduction (CFR), to ensure that the HPA can be operated in its linear regime \cite{spano2017symbol}; then the model in \eqref{eq:non_linear} is approximated as in \eqref{eq:gen_prec}. The other approach is to generalize the linear precoding to non-linear signal pre-distortion (SPD) and devise an iterative approach to obtain the components of SPD. In \cite{mengali2016joint}, this approach is considered along with CFR and is shown to yield performance benefits. The approach opens up a framework to
combine predistortion and precoding.
\section{Trends in Precoding}
\label{sec:trends}
Precoding for current and next generation of satellites has reached a level of academic maturity and industry acceptance. However, as the number of users/ traffic types increase and become  dynamic (spatially, temporally), satellite systems need unprecedented flexibility to adapt to the requirements with the given resources.  Thus the future generation of satellites would differ in (i) the amount of flexibility needed to adapt to the offered services  and (ii) the payload processing to offer this flexibility \cite{Payload}. These aspects are briefly discussed next.
\subsection{Flexible precoding}
The evolution of the satellite system architecture evidence a trend towards flexible, reconfigurable and cost-efficient payloads, motivated by the non-uniform traffic demand and powered by the advances in Software-Defined Radio  technology \cite{RRM, Payload}. The benefits of precoding techniques over broadband high-throughput satellite systems with some kind of flexibility built into them has so far received limited attention from the research community. Flexibility can be implemented mainly in two forms \cite{RRM}: (i) flexible allocation of
bandwidth \cite{Tedros2020}; or (ii) time flexibility (beam hopping) \cite{Lei2020}. In \cite{Tedros2020}, the limits of a frequency-flexible GEO satellite system without precoding capabilities are explored in terms of achievable user demand satisfaction rate. The combination of precoding and beam hopping is investigated in \cite{Kibria2019}, where the individual competencies of both techniques are seamlessly combined resulting into the novel cluster hopping concept. However, these works represent preliminary studies which need to be further developed in order to ensure a success of the precoding technology.
\subsection{Payload processing: Hybrid precoding}
It is envisaged that the next generation systems support a large number of beams through an elaborate feed configuration. The underlying large antenna structures can be implemented on the spacecraft thanks to the migration to mmWave frequencies and development of compact antenna designs. However, the dimension of the precoder matrix increases with the number antennas (feeds). This causes a significant digital processing overload, necessitating its implementation on-ground to reduce processing complexity on board. However, RTT limits the impact of flexibility when processing on-ground. In this context, payloads supporting hybrid precoding comprising processing in analog and digital domains are being promoted enable on-board implementations with low complexity and negligible loss in performance.
%, while heralding the exploitation of on-board processing. 

The hybrid architecture involves synthesizing a precoder as a cascade of a lower dimensional digital precoding followed by a analogue processing implementing a network of phase shifts. Recalling the definition of ${\bf W}$ from \eqref{eq:Prec_def}. the idea is to approximate ${\bf W}\approx {\bf F}_{RF} {\bf F}_{BB}$, where ${\bf F}_{RF}$ is the $N_f \times N_{rf} $ analog beamformer and ${\bf F}_{BB}$ is the $N_{rf} \times N_b$ digital implementation with $N_b < N_{rf}<< N_f$. This decomposition  reduces processing complexity, power consumption and the hardware cost. Several works have  pursued the optimal design of ${\bf F}_{RF}, {\bf F}_{BB}$ for a given ${\bf W}$ and system requirements (e.g. number of RF chains). The desired properties of these entities (e.g, output power, phase-only etc.) are included as constraints; kindly refer to \cite{Aakash} for recent results. 
%, 

{Precoding and Beamforming:} The hybrid processing design mentioned above is attractive in combining precoding and beamforming (cf. Section \ref{ssec:Prec_BF}). In particular, if ${\bf F}_{BB}$ is designed to adapt to CSI and ${\bf F}_{RF}$ is designed considering macro-aspects like traffic evolution and coverage, then, ${\bf F}_{RF}$ is simply an {\em analog beamformer} \cite{8969518}, while ${\bf F}_{BB}$ is the {\rm precoder}. Such a design is novel having drawn limited attention till date.
\vspace*{-0.2in}
\section{Conclusions}
Precoding in satellite systems has lagged its terrestrial counterpart. However, the nuances of the satellite systems do not allow for the mere application of existing code-book based precoding methods e.g. LTE-A.  The paper presents a canvas of the precoding techniques for the current  satellite  systems and the path envisaged for the future generation flexible satellites. 
%
%\bibliographystyle{IEEEtran}
%\bibliography{references}

\end{document}